\documentclass[twocolumn,showpacs,preprint]{revtex4}
\usepackage{amsfonts}
\usepackage{amsmath}
\usepackage{graphicx}
\begin{document}
\draft
\title{Superconducting critical fields of K$_{0.8}$Fe$_{2}$Se$_2$ single crystal}
\author{M.I. Tsindlekht$^1$,  I. Felner$^1$, M. Zhang$^2$, A. F. Wang$^2$  and X. H. Chen$^2$}
\affiliation{$^1$The Racah Institute of Physics, The
Hebrew University of Jerusalem, 91904 Jerusalem,
Israel}
\affiliation{$^2$Hefei National Laboratory for Physical Science at Microscale and Department of Physics,
University of Science and Technology of China, Hefei, Anhui 230026, People's Republic of China}

\begin{abstract}

We report the results of an experimental study of dc and low frequencies magnetic properties of K$_{0.8}$Fe$_{2}$Se$_2$ single crystal when the dc magnetic field is applied parallel to the $\bf{ab}$ plane. From the data obtained, we deduce the full H-T phase diagram which consists of all three H$_{c1}$(T), H$_{c2}(T)$ and H$_{c3}(T)$ critical magnetic field plots. The two H$_{c1}$(T) and  H$_{c2}$(T) curves were obtained from dc magnetic measurements, whereas the surface critical field H$_{c3}$(T) line was extracted by ac susceptibility studies. It appears that near T$_c$, the H$_{c3}$(T)/H$_{c2}$(T) ratio is $\approx 4.4$ which is much larger than expected.

\end{abstract}
\date{\today}
\maketitle


Over the last four decades the ternary intermetalic compounds, which crystallize in the body-centered tetragonal ThCr$_2$Si$_2$ (space group \textit{I4/mmm}), have been of great interest due to the variety of physical phenomena observed in these materials. As early as 1973, both magnetic and M\"{o}ssbauer (MS) effect spectroscopy studies suggested that in RFe$_2$M$_2$ (R=rare-earth, M=Si or Ge), the Fe ions are diamagnetic~\cite{1}. Indeed, neutron powder diffraction measurements on NdFe$_2$Si$_2$ confirmed the absence of any magnetic moment on the Fe sites, and determined that the Nd sublattice is antiferromagnetically (AFM) ordered at T$_N\approx 16$ K, with the moments aligned along the $\bf{c}$ axis~\cite{2}.

At high temperatures both BaFe$_2$As$_2$ and MFe$_2$Se$_2$ (M=K ,Rb Cs, Tl/K and Tl/Rb) pristine materials also crystallize in this tetragonal ThCr$_2$Si$_2$ type structure. The common properties of these systems are that the Fe-As and Fe-Se layers exhibit long-range three-dimensional AFM order at T$_N\approx $140-150 K and $\approx 520-550$ K, with Fe$^{2+}$ moment of 0.87(3) $\mu_B$/Fe or 3.3 $\mu_B$/Fe respectively. In BaFe$_2$As$_2$ the Fe moments are aligned within the $\bf{ab}$ plane~\cite{3} whereas in MFe$_2$Se$_2$ they are along the $\bf{c}$-axis~\cite{4}. Also associated with or preceding the magnetic transition is a structural transition: tetragonal to orthorhombic for the pnictides, and tetragonal (\textit{I4/mmm}) to another tetragonal (\textit{I4/m}) structure for the Fe-Se based materials. The major difference between the two systems, noticeable from several types of measurements, is that in the Fe-As based materials the temperature composition complex phase diagrams show a generic behavior as a function of the substituent concentration (\textit{x}). This implies a systematic suppression of the magnetic transition by increasing \textit{x} by either electrons or holes~\cite{5}. Then, above a critical concentration (which depends on the substituent), SC is observed. Indeed, partial substitution of Ni or Co for Fe in BaFe$_2$As$_2$ induces superconductivity (SC) in the Ba(Fe$_{1-x}$Ni$_x$)$_2$As$_2$ and Ba(Fe$_{1-x}$Co$_x$)$_2$As$_2$ systems~\cite{5,6}. On the other hand, the non-stoichiometric M$_x$Fe$_{2-x}$Se$_2$ materials also become SC around 30-33 K, but the AFM state persists even at low temperatures. That means that in M$_x$Fe$_{2-x}$Se$_2$ a real coexistence of the two states occurs, since both states are confined to the same Fe-Se crystallographic layer~\cite{7}. This peculiar property marks such a system as a very  unique one and opens a new avenue for the study of the interplay between magnetism and superconductivity.

So far, the bulk upper critical magnetic field (H$_{c2}$) for SC M$_x$Fe$_{2-x}$Se$_2$ single crystals have been determined over a wide range of temperatures and magnetic fields. For K$_{0.8}$Fe$_{1.76}$Se$_2$, the field dependence of the resistivity at low dc fields (H$_0$) and the radio frequency penetration depth in pulsed magnetic up to 60 T, exhibit an anisotropy in H$_{c2}$ when measured along or perpendicular to the $\bf{c}$-axis~\cite{8}. Generally speaking, a linear  temperature dependence of H$_{c2}$ in both directions is observed and the slope close to T$_c$ is higher for H$_0$ parallel to the $\bf{ab}$ plane. The initial anisotropy factor $\gamma$ is $\approx 2$~\cite{8}. Similar results were obtained for Tl$_{0.58}$ Rb$_{0.42}$Fe$_{1.72}$Se$_2$, where the extrapolation to T=0 yield H$_{c2}$(0) 221T and 44.2 T (an anisotropy $\gamma \approx 5$) for H$_0$ parallel or perpendicular to the $\bf{ab}$ plane respectively~\cite{9,10}.

In this communication, we report on the temperature dependence of the three critical fields H$_{c1}$, H$_{c2}$ and H$_{c3}$ in K$_{0.8}$Fe$_{2}$Se$_2$ single crystal (T$_c\approx 31$ K) measured in a magnetic field applied parallel to the \textit{\bf{ab}} plane. Both H$_{c1}$(T) and H$_{c2}$(T) plots were deduced from dc field isothermal M(H$_0$) curves.  In addition,  the M(H$_0$) curve at 35 K (above T$_c$) is not linear as expected for an AFM material, but rather exhibits a small peculiar hysteresis loop  which is shifted from  the origin, known as the exchange biased field phenomenon. These observations are compared with earlier reports on similar materials. A H$_{c3}$(T) plot was obtained by ac susceptibility measurements. It appears that near T$_c$ H$_{c3}$/H$_{c2}\approx 4.4$, a value which is much higher than the 1.7 predicted for conventional SC material~\cite{PG}.


A single crystal of K$_{0.8}$Fe$_{2}$Se$_2$ was grown by the conventional high-temperature flux method and its actual composition was determined by crystal structure refinement of the powder x-ray diffraction pattern (XRD). Detailed description of the crystal growth procedure and lattice pareameters can be found in Ref~\cite{7}. The size of the triangular-shaped sample is 7 mm (width), 3 mm (height) and 1.5 mm (thickness). The crystal plate is perpendicular to the $\bf{c}$ lattice axis. The temperature and/or field dependence of the dc magnetic moment was measured in a commercial MPMS5 Quantum Design SQUID magnetometer.
Prior to recording the zero-field-cooled (ZFC) curves, the SQUID magnetometer was always adjusted to be in "true" H$_0$ = 0 state. The ac susceptibility $\chi^{\prime}$ and $\chi^{\prime\prime}$ was measured with the pick-up coil method~\cite{SH} at amplitude h$_0$ =0.05 Oe and frequency 1465 Hz. The sample was inserted into one coil of a balanced pair. The amplitude and phase of the unbalanced signal were measured by a lock-in amplifier in a point-by-point mode. The "home-made" measurement cell of the experimental setup was adapted to the SQUID magnetometer.  The block-diagram of this setup was published elsewhere~\cite{LEV2}.

\begin{figure}
\begin{center}
\leavevmode
\includegraphics[width=0.9\linewidth]{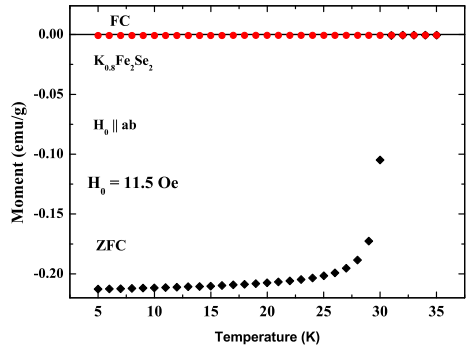}

\caption{(Color online)  ZFC and FC plots measured at 11.5 Oe. }
\label{F1F}
\end{center}
\end{figure}

\textit{(i) M(T) curves.} The ZFC and field-cooled (FC) magnetization curves of  K$_{0.8}$Fe$_{2}$Se$_2$  were measured at 11.5 Oe and are depicted in Fig.~\ref{F1F}. The ZFC branch shows a diamagnetic transition at T$_c$ = 31.0 $\pm 0.5$ K. This transition is not sharp (as expected for a single crystal SC) and it is similar to inhomogeneous materials with a distribution of T$_c$ related to the small spread of stoichiometry inside the sample. The estimated shielding fraction  is about -1/4$\pi$ emu/cc.

\begin{figure}
\begin{center}
\leavevmode
\includegraphics[width=0.9\linewidth]{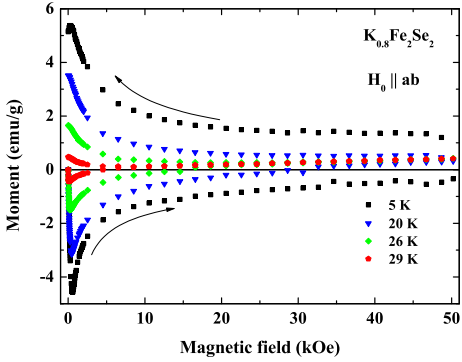}

\caption{(Color online) Ascending and descending branches of M(H$_0$) curves at various temperatures.}
\label{F2F}
\end{center}
\end{figure}

\begin{figure}
\begin{center}
\leavevmode
\includegraphics[width=0.9\linewidth]{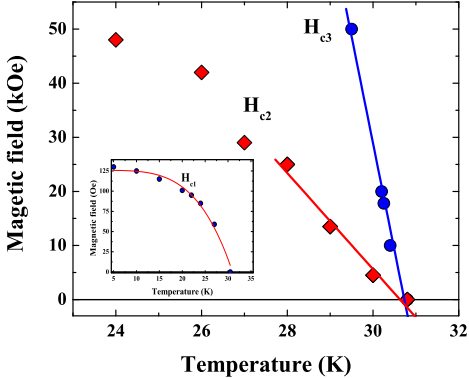}

\caption{(Color online) H-T phase diagram. Inset: Temperature dependence of H$_{c1}$.}
\label{f-1}
\end{center}
\end{figure}

\textit{(ii)  Isothermal M(H$_0$) curves.} Isothermal magnetization curves have been measured at various temperatures and selected M(H$_0$) plots measured at 5, 20, 26 and 29 K are shown in Fig.~\ref{F2F}.
In the Meissner state, the M(H$_0$) curves are linear and H$_{c1}$(T) is defined as the field in which M(H$_0$) deviates from linearity. Temperature dependence of H$_{c1}$ is plotted in Fig.~\ref{f-1} (inset).
H$_{c1}$(T) decreases with T and to obtain the lower critical field at T=0 we used the conventional relation H$_{c1}$(T) = H$_{c1}$(0)(1-(T/T$_{c}$)$^{\alpha}$) which yields:  H$_{c1}$(0) $\approx 130 (5)$ Oe and $\alpha \approx 4.3$. This $\alpha$ is much higher than that expected for conventional BCS materials.

The criterion for determining the upper critical field H$_{c2}$(T) requires consistency, and no one method is entirely unambiguous. The H$_{c2}$(T) values were obtained by two methods: (i) M(T) plots were measured under various applied field (not shown) and H$_{c2}$(T) was defined as the disappearance of the negative signal of the ZFC branches, see Fig.~\ref{F1F}. (ii) The H$_{c2}$(T) data were extracted from Fig.~\ref{F2F}, as the fields where the irreversibility disappears. Both methods yield basically the same results presented in Fig.~\ref{f-1}.
By using the well known Werthamer-Helfand-Honenberg  (WHH) formula~\cite{12}:
$H_{c2}(0) = -0.69T_c(dH_{c2}/dT)$,
where, T$_c\approx 31$ K and the linear slope (close to T$_c$) dH$_{c2}$/dT is -9 kOe/K, Fig.~\ref{f-1}. H$_{c2}$(0) obtained is 193$\pm$6 kOe, a value which is an order of magnitude smaller than that estimated for the same orientation in Tl$_{0.58}$Rb$_{0.42}$Fe$_{1.72}$Se$_2$~\cite{7} and K$_{0.8}$Fe$_{1.8}$Se$_2$~\cite{8}. It is well accepted that the WHH formula is valid for one-band superconductors and that H$_{c2}$(0) might be effected by the complicated multiband structure as observed in various Fe-Se crystals~\cite{ZHA}. Hence, this H$_{c2}$(0) is just a rough estimation. An accurate value can only be achieved by applying high enough magnetic fields.

Using the Ginzburg-Landau (GL) relation for the coherence length: $\xi = (\Phi_0/2\pi  H_{c2})^{1/2}$, we obtained for K$_{0.8}$Fe$_{2-y}$Se$_2$ $\xi(0) \approx 10~nm $.  In order to estimate the second characteristic length, namely the penetration depth $\lambda  (0)$, we use the useful relationship:
$2H_{c1}(0)/ H_{c2}(0) = (\ln(\kappa)+0.5)/\kappa^2$,
where $\kappa= \lambda(0)/\xi  (0)$ is the GL parameter. Solving this equation numerically we obtained $\kappa\approx 58$ from which  $\lambda (0) \approx 580$ nm is deduced.

From the experimental hysteresis loop width the critical current density (J$_c)$ can be estimated by using the Bean critical state model: $J_c = 20\Delta M/a(1-a/3b)$,  where $\Delta M$ is the magnetization loop width at a given H$_0$, and \textit{a} and \textit{b} are the crystal dimensions perpendicular to H$_0$ where \textit{a}$ <$\textit{b}. At 5 K for H$_0=10$ kOe the estimated J$_c$ is $\approx 10^3$ A/cm$^2$, a value which does not change much with field (up to 50 kOe) and agrees well with Ref.~\cite{13,16}. Of course, this estimate value is very crude, but shows that the current sample cannot support large critical currents even at low temperatures.

\begin{figure}
\begin{center}
\leavevmode
\includegraphics[width=0.9\linewidth]{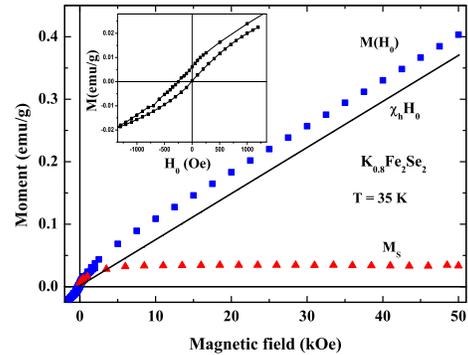}

\caption{(Color online)  Magnetization curves in a normal state, T = 35 K. The inset shows the shifted hysteresis loop obtained at low H$_0$. See text.}
\label{F4F}
\end{center}
\end{figure}

At 35 K (above T$_c$)  at low H$_0$, the M(H$_0$) curve is not linear as expected for an AFM material (see Ref.~\cite{13}) and may contain another minor ferromagnetic (FM) component as an extra phase not detectable by XRD (Fig.~\ref{F4F}). At high H$_0$ M(H$_0$) is composed of a linear ($\chi_h \times H_0$) and a saturation M$_S$ term. In accordance with the estimation, $\chi_h\approx 7.4\times 10^{-6}$ emu/$g\times Oe$. Subtracting the linear part yields the saturation moment M$_S$ = 0.033 emu/g, which is attributed to 0.015 \% of pure iron. Indeed, the presence of pure Fe was confirmed by our $^{57}$Fe MS studies performed on the same crystal, to be published elsewhere. Irreversibility in M(H$_0$) is observed at low H$_0$, and the hysteresis loop obtained is shown in Fig.~\ref{F4F} (inset).
This loop is not symmetric relative to the origin and is known as the exchange-bias (EB) phenomena~\cite{14,15}. EB is associated with the exchange anisotropy created at the interface between an AFM and a FM materials. The main information deduced from this hysteresis loop is: the magnetic coercive field  H$_c$ = 130 Oe and exchange bias E$_x$ =65 Oe which is the loop shift from the origin. The discussion on EB is beyond the scope of the present paper. However we may intuitively assume that this phenomenon is caused by the  FM Fe particles immersed in the AFM matrix.

\begin{figure}
\begin{center}
\leavevmode
\includegraphics[width=0.9\linewidth]{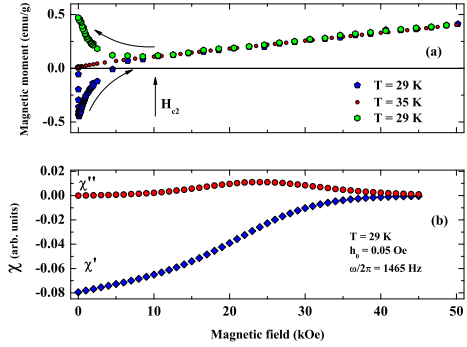}

\caption{(Color online) Panel \textit{a}: Magnetization curve at T= 29 K and 35 K. Panel \textit{b}: Field dependencies of $\chi^{\prime}$ and $\chi^{\prime\prime}$ at T = 29 K.}
\label{f-1a}
\end{center}
\end{figure}

\begin{figure}
\begin{center}
\leavevmode
\includegraphics[width=0.9\linewidth]{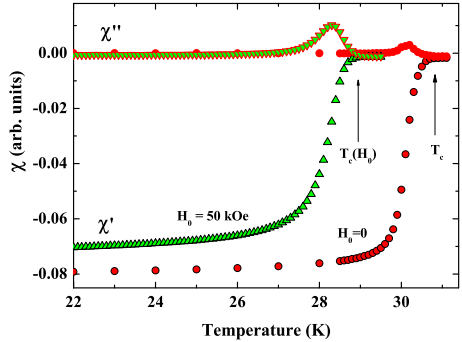}

\caption{(Color online) Temperature dependencies of $\chi^{\prime}$ and $\chi^{\prime\prime}$ at H$_0$ = 0 and 50 kOe. }
\label{f-3}
\end{center}
\end{figure}

It is well accepted that ac susceptibility studies are a powerful tool for determining of the surface superconducting states (SSS) including determination of the surface critical field H$_{c3}$(T), see~\cite{JUR,SHT} and references therein. A comparison between dc M(H$_0$) and ac susceptibility measurements is depicted in Fig.~\ref{f-1a}. Fig.~\ref{f-1a}\textit{a} shows the ascending and descending M(H$_0$) curves measured at 29 K and ascending M(H$_0$) curve at 35 K from which H$_{c2}$ can easily determined (with the accuracy of $\pm 1$ kOe) as discussed above. Note that the plot measured at 35 K, coincides with the reversible data collected at 29 K. On the other hand, the real $\chi^{\prime}$ and the imaginary $\chi^{\prime\prime}$ ac susceptibility plots, measured under the same conditions, demonstrate clearly the existence of SC up to H$_{c3}> 50$ kOe which is well above H$_{c2}\approx 13$ kOe. Two more examples of determining H$_{c3}$ from $\chi^{\prime}$(T) and $\chi^{\prime\prime}$(T) measured at H$_0$ = 0 and 50 kOe are presented in Fig.~\ref{f-3}. Here we can obtain T$_c(H_0)$ with an accuracy about $\pm 0.5$ K. From the obtained data H$_{c3}$(T) was deduced and shown in Fig.~\ref{f-1}. Note that at H$_0$ = 0 the value obtained is T$_c$. The slope of the H$_{c3}$(T) curve near T$_c$ is $\approx -40$ kOe/K. Therefore the H$_{c3}$/H$_{c2}$ ratio is $\approx 4.4$, a value which is much larger than the 1.7 predicted for single band conventional superconductors.
In several publications, H$_{c2}$(T) was deduced from resistivity and/or ac susceptibility measurements. Indeed, these studies provide accurate H$_{c2}$(T) values when the dc field is applied perpendicular to the $\bf{ab}$ crystal plane. In this geometry, the nucleation of the SC state starts at H$_0<$ H$_{c2}$~\cite{PG}. On the other hand, for $\overrightarrow{H_0}$ parallel to $\bf{ab}$ plane, this nucleation starts at H$_{c3}$ which is always higher than H$_{c2}$~\cite{PG,JUR,SHT}. For this geometry, H$_{c2}$(T) can be determined from bulk measurements such as dc M(H$_0$) (see Fig.~\ref{f-1a}\textit{a}) and/or specific heat capacitance studies. Therefore the high H$_{c2}$(T) values reported for $\overrightarrow{H_0}$ parallel to $\bf{ab}$ plane in Ref.~\cite{7,8} and in several other publications, are presumably  the surface H$_{c3}$(T) plots. This issue needs more consideration.

In summary, we have performed dc and ac magnetization measurements from which all three critical fields for the SC K$_{0.8}$Fe$_{2}$Se$_2$  single crystal are determined. Evaluating H$_{c1}(0)\approx 130$ Oe and H$_{c2}(0)\approx 193$ kOe permits us to calculate the coherence length $\xi (0)\approx 10$ nm and the penetration depth $\lambda  (0)\approx 580$ nm. Our ac susceptibility study provides for the first time the determination of H$_{c3}(T)$ for the $\overrightarrow{H_0}$ parallel to $\bf{ab}$ plane. The high H$_{c3}/ H_{c2}\approx 4.4$ obtained needs more consideration.

The research in Jerusalem is partially supported by the Israel Science Foundation (ISF, Bikura 459/09), by the joint German-Israeli DIP project and by the Klachky Foundation for Superconductivity. The authors are deeply grateful to V.M. Genkin for valuable discussions.

\end{document}